\documentclass[aps,pra,floatfix,superscriptaddress,twocolumn,showpacs,showkeys,10pt]{revtex4-1}
\usepackage{amssymb, amsmath,color,mciteplus,graphicx,subfigure}
\usepackage{calc,epsfig,epstopdf,color,mciteplus,bm,mathrsfs,multirow}
\usepackage{times}
\usepackage{float}
\usepackage{lipsum}

\begin{document}

\title{Composite nonadiabatic geometric quantum gates with optimization on superconducting circuits}

\author{Cheng-Yun Ding}
\affiliation{School of Mathematics and Physics, Anqing Normal University, Anqing 246133, China}

\author{Wan-Fang Liu}
\affiliation{School of Mathematics and Physics, Anqing Normal University, Anqing 246133, China}

\author{Li-Hua Zhang}
\affiliation{School of Electronic Engineering and Intelligent Manufacturing, Anqing Normal University,
Anqing 246133, China}

\author{Jian Zhou}\email{jianzhou8627@163.com}
\affiliation{School of Electrical and Photoelectronic Engineering, West Anhui University, Luan 237012, China}

\author{Zheng-Yuan Xue}\email{zyxue83@163.com}
\affiliation{Key Laboratory of Atomic and Subatomic Structure and Quantum Control (Ministry of Education),\\
Guangdong Basic Research Center of Excellence for Structure and Fundamental Interactions of Matter,\\
and School of Physics, South China Normal University, Guangzhou 510006, China}
\affiliation{Guangdong Provincial Key Laboratory of Quantum Engineering and Quantum Materials,\\
Guangdong-Hong Kong Joint Laboratory of Quantum Matter, and Frontier Research Institute for Physics,\\
South China Normal University, Guangzhou 510006, China}

\date{\today}

\begin{abstract}
Due to its fast and robust characteristics, nonadiabatic geometric quantum computation with various optimized techniques has received much attention. However, these strategies either require precise pulse control or can only mitigate partial systematic errors, hindering their experimental development. Here, we propose a scheme for optimized composite nonadiabatic geometric quantum gates (OCNGQGs), which can further enhance the gate performance of the composite nonadiabatic geometric scheme. Specifically, by optimizing the path parameter, our scheme effectively resists systematic errors in both directions, i.e., Rabi frequency and detuning errors, while preserving the flexibility of pulse shapes. Numerical simulations demonstrate that our scheme offers superior gate robustness against these two types of errors compared to conventional schemes. Additionally, we propose to implement our scheme on superconducting transmon qubits, where the numerical results show the robustness of universal gates remaining evident within current technology. Therefore, our proposal provides a promising approach to achieve robust quantum gates  for future scalable quantum computation.
\end{abstract}

\maketitle
\section{Introduction}
Using quantum error correction codes to achieve logical quantum gates with lower errors is a crucial step in the realization of universal quantum computing, and it has been experimentally achieved the break-even point \cite{sivak2023real,ni2023beating}. However, higher quantum gate-fidelity is more desirable, as it needs fewer physical qubits for error correction, significantly reducing resource requirements, especially in the current era. Therefore, achieving high-fidelity quantum gates on physical qubits is essential for the implementation of scalable universal quantum computing.

During the past decades, geometric quantum computing (GQC) has made significant strides as a robust computational strategy, both theoretically \cite{ekert2000geometric,zhang2023geometric,liang2023nonadiabatic} and experimentally \cite{jones2000geometric,abdumalikov2013experimental,zu2014experimental,huang2019experimental,xu2020experimental,zhao2021experimental,xu2021demonstration}. Its use of geometric phases \cite{berry1984quantal,aharonov1987phase} to construct quantum gates inherently resists local operation errors, as these phases depend only on the global properties of the evolution path rather than details of the evolution. Early GQC relied on adiabatic geometric phases, such as the Berry phase \cite{berry1984quantal}, which required the system to evolve slowly enough to ensure that the evolution states remain in certain instantaneous energy eigenstates without undergoing mutual transitions, exacerbating the impact of environmental noise on the quantum gate. However, slow evolution will lead to more decoherence-induced errors. To address this, nonadiabatic GQC (NGQC), based on nonadiabatic geometric phases \cite{aharonov1987phase}, has been introduced \cite{xiang2001nonadiabatic,zhu2002implementation}, significantly accelerating quantum gate operations.

In particular, the NGQC scheme based on the orange-slice-shaped single-loop evolution path has been theoretically proposed \cite{zhao2017rydberg,chen2018nonadiabatic} and experimentally verified \cite{xu2020experimental,zhao2021experimental}, which further reduces the gate operation time. Subsequently, to enhance gate performance beyond dynamical quantum gates, various optimization schemes have been proposed, including the combination of various techniques such as composite pulses \cite{zhu2019single,zhou2021nonadiabatic,liang2022composite,fang2024nonadiabatic}, dynamical correction\cite{li2021dynamically,liang2022robust,ding2023dynamical}, path optimization \cite{li2020approach,ding2021path,chen2024universal,liang2024nonadiabatic,lv2020noncyclic,ma2023noncyclic}), shortcut to adiabaticity \cite{qiu2021experimental,li2021noncyclic}, optimal control \cite{zhang2019searching,xu2020nonadiabatic,liang2024nonadiabatic1,zhang2025multiobjective}, dynamical decoupling \cite{wu2022nonadiabatic}, etc. Most of these schemes can only resist one of the two typical types of errors, i.e., Rabi frequency and detuning errors, denoted as X and Z errors, rather than both. Although the globally robust scheme \cite{liang2024nonadiabatic1} can be achieved by finely controlling the pulse waveform, complex pulse waveforms are not user-friendly in experiments. Note
that, conventional composite nonadiabatic geometric quantum gates (CNGQGs) \cite{chen2018nonadiabatic,zhou2021nonadiabatic} have no restrictions on the pulse waveform for given amplitude peak, and it has two different robust path configurations, namely ``Path A" and ``Path B", which can resist X and Z errors, respectively \cite{zhu2019single,zhou2021nonadiabatic}. That is, a certain path configuration can resist errors in a certain direction but exhibit the converse effect on the other one.

To overcome this dilemma, we here propose a scheme for optimized CNGQGs (OCNGQGs). By optimizing the path parameter, our scheme can not only achieve stronger gate robustness against errors in a particular direction, but also realize nonadiabatic geometric gates that can simultaneously resist errors in both directions. Besides, we propose to implement our scheme on superconducting transmon qubits. Numerically simulation of the gate performance demonstrates the significant advantages of our scheme compared with others. Thus, our scheme provides a promising way to implement more robust quantum gates, which is essential for future scalable quantum computation.


\begin{figure}[tbp]
\centering
\includegraphics[width=1.0\linewidth]{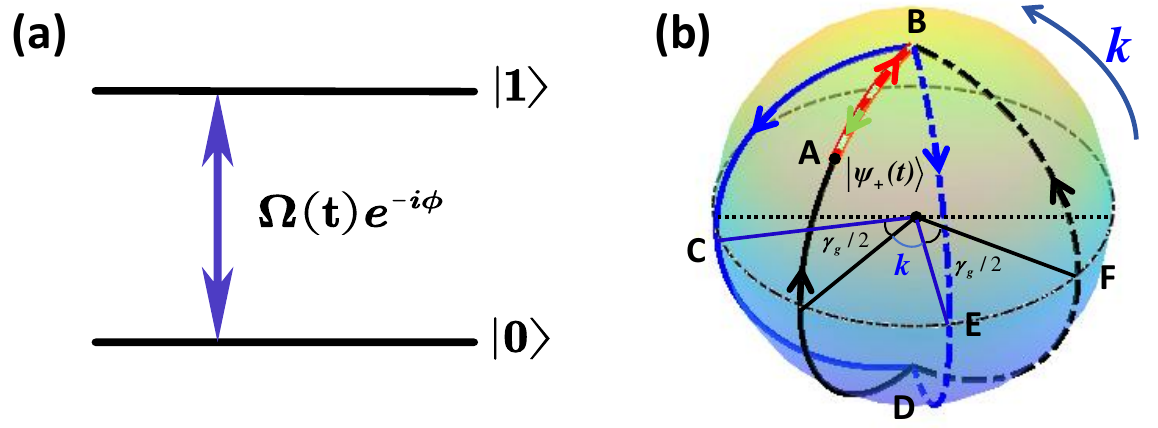}
\caption{Illustration of our scheme. (a) An ideal two-level quantum system, resonantly driven by a microwave field, wherein $\Omega(t)$ and $\phi$ denote the respective driving strength and phase. (b) The closed evolution trajectory, designated as ``Path A", for our universal single-qubit quantum gates on the Bloch sphere, which completes a cyclic evolution from point $\textrm{A}$ along $\textrm{AB}-\textrm{BCD}-\textrm{DA}-\textrm{AB}-\textrm{BED}-\textrm{DFB}-\textrm{BA}$ back to the starting point, and $k$ signifies the angle formed between the two orange-petal-shaped paths.}\label{Figure1}
\end{figure}

\section{Optimized composite nonadiabatic geometric quantum gates}\label{sec2}

In this section, we first present our optimized scheme for NGQGs, and then compare the gate performance of our scheme with other ones.

\subsection{The gate construction}
Consider an ideal two-level system encoded as computational bases $|0\rangle$ and $|1\rangle$, as shown in Fig. \ref{Figure1}(a), then the resonant Hamiltonian driven by an external microwave field, under the rotating framework, can be expressed as
\begin{equation}
\mathcal{H}_1(t)=\frac{\Omega(t)}{2}\left(\cos\phi\sigma_x+\sin\phi\sigma_y  \right),
\end{equation}
where $\Omega(t)$ is the Rabi frequency and $\phi$ is the phase of driving microwave field, and $\sigma_x$ and $\sigma_y$ are well-known Pauli operators.
Consider a set of orthogonal dressed-states of system $|\psi_\pm(t)\rangle$ as evolution states, expressed as
\begin{eqnarray}
\begin{split}
|\psi_+(t)\rangle&=e^{-if(t)}\left[\cos\frac{\alpha(t)}{2}|0\rangle+\sin\frac{\alpha(t)}{2}e^{i\beta(t)}|1\rangle\right], \\
|\psi_-(t)\rangle&=e^{if(t)}\left[\sin\frac{\alpha(t)}{2}e^{-i\beta(t)}|0\rangle-\cos\frac{\alpha(t)}{2}|1\rangle\right],
\end{split}
\end{eqnarray}
whose corresponding evolution path on the Bloch sphere can be determined by the time-dependent polar angle $\alpha(t)$ and azimuthal angle $\beta(t)$. And, $f(t)$ is the global phase, which cannot be visualized on the Bloch sphere.

According to the Schr\"{o}dinger equation $i|\dot{\psi}_\pm(t)\rangle=\mathcal{H}_1(t)|\psi_\pm(t)\rangle$, the relationship between the Hamiltonian parameters $\Omega(t)$, $\phi$ and the parameters $\alpha(t)$, $\beta(t)$ of the evolution path can be obtained as follows
\begin{eqnarray}
\begin{split}
\dot{\alpha}(t)&=\Omega(t)\sin[\phi-\beta(t)], \\
\dot{\beta}(t)&=-\Omega(t)\cot\alpha(t)\cos[\phi-\beta(t)].
\end{split}
\end{eqnarray}
Based on this, we can solve the Hamiltonian parameters in reverse according to the path parameters, that is, we can choose an appropriate evolution path to reversely engineer the Hamiltonian parameters meeting specific conditions. Consequently, when the evolution states complete any cyclic evolution with a period of time $\tau$, on the Bloch sphere, they will become $|\psi_\pm(\tau)\rangle=U(\tau)|\psi_\pm(0)\rangle=e^{\mp if(\tau)}|\psi_\pm(0)\rangle$, and the corresponding evolution operator can be calculated as
\begin{equation}
U(\tau)=\sum_{j=\pm}e^{\mp if(\tau)}|\psi_j(0)\rangle\langle\psi_j(0)|=e^{-if(\tau)\vec{n}\cdot\vec{\sigma}},
\end{equation}
in which $f(\tau)$ is the accumulated total phase during the entire evolution process, with an initial value of $0$ and its dynamics cannot be visualized on the Bloch sphere. Note that, $f(\tau)$ includes two components, i.e., the dynamical and geometric phases. They can be calculated, respectively, as
\begin{eqnarray}
\begin{split}
\gamma_d&=-\int_0^{\tau}\langle\psi_+(t)|\mathcal{H}_1(t)|\psi_+(t)\rangle dt \\
&=\frac{1}{2}\int_{0}^{\tau}\dot{\beta}(t)\sin{\alpha(t)}\tan{\alpha(t)}dt,  \\
\gamma_g&=i\int_0^{\tau}\langle\psi_+(t)|\frac{\partial}{\partial t}|\psi_+(t)\rangle dt \\
&=-\frac{1}{2}\int_{0}^{\tau}\dot{\beta}(t)[1-\cos{\alpha(t)}]dt.
\end{split}
\end{eqnarray}
Furthermore, $\vec{n}=(\sin\alpha_0\cos\beta_0,\sin\alpha_0\sin\beta_0,\cos\alpha_0)$ is a unit vector for any direction with $\alpha_0$ and $\beta_0$ being the initial polar and azimuth angles respectively, and $\vec{\sigma}=(\sigma_x,\sigma_y,\sigma_z)$ is the unit Pauli vector. Thus, $U(\tau)$ denotes a single-qubit gate that rotates around an arbitrary axis $\vec{n}$ by an angle of $2f(\tau)$.

Here, to obtain quantum gates with pure geometric phases, we choose a composite evolution path, which is divided into seven stages. That is, the first three stages are that the evolution state $|\psi_+(t)\rangle$ starts from the initial point $\textrm{A}$ in the Bloch sphere, and then followed the meridian $\textrm{AB}$ to the North pole $\textrm{B}$, and followed another meridian $\textrm{BCD}$ to South pole $\textrm{D}$ and finally run back the starting point along the meridian $\textrm{DA}$; the last four ones are followed the meridians $\textrm{AB}$, $\textrm{BED}$, $\textrm{DFB}$ and $\textrm{BA}$, respectively, i.e., the entire evolution path is $\textrm{AB}-\textrm{BCD}-\textrm{DA}-\textrm{AB}-\textrm{BED}-\textrm{DFB}-\textrm{BA}$, as shown in Fig. 1(b). It can be seen that both orange-sliced paths have an inner angle of $\gamma_g/2$, which can also be set to $\pi-\gamma_g/2$. Therefore, the former is called ``Path A", and the latter as ``Path B", detailed discussion of the path configurations is presented in Appendix A. According to the reverse design above, the Hamiltonian parameters of the system are correspondingly set as
\begin{eqnarray}
\begin{split}
\int_0^{\tau_1}\Omega(t)dt&=\alpha_0, \phi_1=\beta_0-\frac{\pi}{2}, t\in[0,\tau_1],\\
\int_{\tau_1}^{\tau_2}\Omega(t)dt&=\pi, \phi_2=\beta_0-\frac{\gamma_g}{2}+\frac{\pi}{2},t\in[\tau_1,\tau_2],\\
\int_{\tau_2}^{\tau_3}\Omega(t)dt&=\pi-\alpha_0, \phi_3=\beta_0-\frac{\pi}{2},t\in[\tau_2,\tau_3], \\
\int_{\tau_3}^{\tau_4}\Omega(t)dt&=\alpha_0, \phi_4=\beta_0-\frac{\pi}{2}, t\in[\tau_3,\tau_4],\\
\int_{\tau_4}^{\tau_5}\Omega(t)dt&=\pi, \phi_5=\beta_0+k+\frac{\pi}{2}, t\in[\tau_4,\tau_5],\\
\int_{\tau_5}^{\tau_6}\Omega(t)dt&=\pi, \phi_6=\beta_0+k+\frac{\gamma_g}{2}-\frac{\pi}{2}, t\in[\tau_5,\tau_6],\\
\int_{\tau_6}^{\tau}\Omega(t)dt&=\alpha_0, \phi_7=\beta_0+\frac{\pi}{2}, t\in[\tau_6,\tau] .
\end{split}
\end{eqnarray}
After completing a cyclic evolution with period $\tau$, the evolution operator can be calculated as
\begin{eqnarray}
\begin{split}
U(\tau)&=U(\tau,\tau_6)U(\tau_6,\tau_5)U(\tau_5,\tau_4) \\
&\quad\cdot U(\tau_4,\tau_3)U(\tau_3,\tau_2)U(\tau_2,\tau_1)U(\tau_1,0) \\
&=\cos\gamma_g\,I-i\sin\gamma_g\left(
                              \begin{array}{cc}
                                \cos\alpha_0 & \sin\alpha_0e^{-i\beta_0} \\
                                \sin\alpha_0e^{i\beta_0} & -\cos\alpha_0 \\
                              \end{array}
                            \right), \\
&=e^{-i\gamma_g\vec{n}\cdot\vec{\sigma}}
\end{split}
\end{eqnarray}
which can lead to the universal set of single-qubit quantum gates. As the evolution path always follows the meridian, the corresponding dynamical phases remain zero throughout the process, and the remaining phase $\gamma_g$ of $f(\tau)$ is the geometric one and $U(\tau)$ is the geometric gate. Notice that, the path parameter $k$ corresponds to the angle between the $\textrm{DAB}$ and $\textrm{BED}$ meridians, and it can be an arbitrary constant within the interval $[0,2\pi]$. From the Fig. 1(b), we can see that the orange-sliced path consisted of meridians \textrm{BED} and \textrm{DFB} rotates anticlockwise as $k$ increases, and will coincide with the other orange-sliced path consisted of meridians \textrm{BCD} and \textrm{DAB} at $k=2\pi-\gamma_g/2$. Finally, when $k=2\pi$, it can complete a cycle. Clearly, a different selection of the path parameter $k$ will correspond to different composite evolution paths, which usually possess various error sensitivity properties. Thus, it can be adjusted to achieve OCNGQGs with stronger robustness.

To search for the optimal parameter $k$ under noisy condition, we here consider two typical types of errors, i.e., the X and Z errors, to evaluate the infidelity of target quantum gates. Then, the noisy Hamiltonian can be written as
\begin{equation}
\mathcal{H}'_1(t)=-\frac{\delta\Omega_m}{2}\sigma_z+\frac{1+\epsilon}{2}\Omega(t)(\cos\phi\sigma_x+\sin\phi\sigma_y),
\end{equation}
where $\epsilon$ and $\delta$ are the ratios of X and Z errors, respectively, and $\Omega_m$ is the maximum of the $\Omega(t)$. The dynamical process of the quantum system is modeled and simulated by the Lindblad master equation \cite{lindblad1976generators} of
\begin{equation}
\dot{\rho}_1(t)=i[\rho_1(t),\mathcal{H}'_1(t)]+\frac{1}{2}\sum_{m=z,-}\kappa_m\mathcal{L}(\mathcal{A}_m),
\end{equation}
in which $\rho_1(t)$ is the density operator of the total quantum system driven by the noisy Hamiltonian $\mathcal{H}'_1(t)$; $\mathcal{L}(\mathcal{A})=2\mathcal{A}\rho_1\mathcal{A}^{\dagger}-\mathcal{A}^{\dagger}\mathcal{A}\rho_1-\rho_1\mathcal{A}^{\dagger}\mathcal{A}$ is the Lindblad operator for $\mathcal{A}$; $\mathcal{A}_z=|0\rangle\langle0|-|1\rangle\langle1|$; $\mathcal{A}_-=|0\rangle\langle1|$; and $\kappa_z$, $\kappa_-$ are the dephasing and decay rates, respectively.

\begin{figure}[tbp]
\centering
\includegraphics[width=0.9\linewidth]{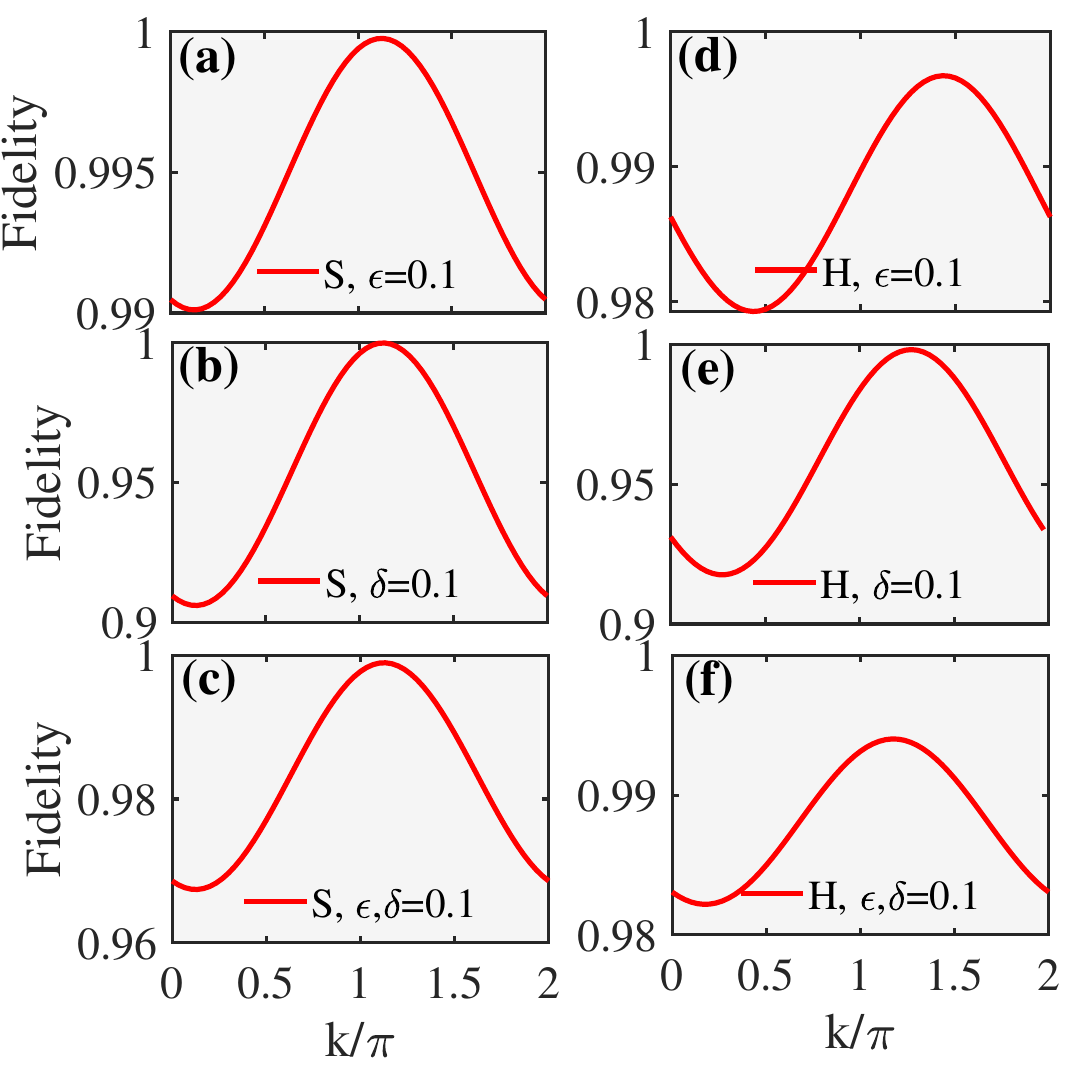}
\caption{The fidelities as a function of the path parameter $k$ against X, Z and both errors, respectively, for optimized composite geometric (a)-(c) S and (d)-(f) H gates, without decoherence.}\label{Figure2}
\end{figure}

In addition, the gate fidelity can be defined as \cite{schwinger1960unitary,1523643}
\begin{equation}
F=\frac{1}{6}\sum_{m=1}^{6}\langle\psi_m(0)|U^{\dagger}(\tau)\rho_1(\tau)U(\tau)|\psi_m(0)\rangle,
\end{equation}
with the initial state $|\psi_m(0)\rangle$ being any element in the set $\{|0\rangle,|1\rangle,(|0\rangle+i|1\rangle)/\sqrt{2},(|0\rangle-i|1\rangle)/\sqrt{2},(|0\rangle+|1\rangle)/\sqrt{2},(|0\rangle-|1\rangle)/\sqrt{2}\}$. These six initial states are sufficient to accurately evaluate any single qubit gate, as they form a set of over-complete bases in two-dimensional space. The influence of quantum gates on any two-dimensional initial state can be decomposed into their effects on these three directional axis states. First, without decoherence errors, i.e., $\kappa_z=\kappa_-=0$, we take the phase (S) and Hadamard (H) gates as two typical examples, corresponding to $(\alpha_0,\beta_0,\gamma_g)=(0,0,\pi/4)$ and $(\alpha_0,\beta_0,\gamma_g)=(\pi/4,0,\pi/2)$, and test the corresponding gate fidelities as a function of the optimized parameter $k$ under these two typical errors, respectively. As shown in Fig. 2, the fidelities of the target S or H gates vary across the parameter interval, indicating different sensitivity to errors for different $k$ values. For instance, for geometric S gate, the gate robustness against these errors is slightly decreased at $k\in[0,0.13\pi]$, and it continues to increase and decrease at $k\in[0.13\pi,1.13\pi]$ and $k\in[1.13\pi,2\pi]$, respectively. The situation for the geometric H gate is similar. This allows us to identify the optimal $k$ to implement more robust geometric gates. Here, the optimal values $k$ are listed in Table I, which are those against the X error, the Z error, and both errors, respectively. When $k=2\pi-\gamma_g/2$, our optimized composite geometric paths will reduce to conventional composite geometric schemes. For example, for the S gate ($\gamma_g=\pi/4$), the path parameter $k=1.875\pi$, while it is $k=1.75\pi$ for the H gate with geometric phase $\gamma_g$ being $\pi/2$. Besides, Fig. 2 also shows these parameters is not optimal settings against various errors, confirming the effectiveness of our optimized scheme for more robust geometric gates. Furthermore, note that the pulse waveform is set as square wave $\Omega(t)=\Omega_m=\Omega$ and $\epsilon=\delta=0.1$ for demonstration purposes in Fig. 2.

\setlength{\tabcolsep}{5.5mm}{\begin{table}
\centering
\caption{The optimal parameters $k$ for our geometric S and H gates against X, Z and both errors, respectively.}
\begin{tabular}{cccc}
\hline\hline\noalign{\smallskip}
Gates$\diagdown$ Errors & X&Z& X\&Z \\
\noalign{\smallskip}\hline\noalign{\smallskip}
$\textrm{S}$ & $1.13\pi$ & $1.13\pi$ & $1.13\pi$\\
$\textrm{H}$ & $1.43\pi$ & $1.27\pi$& $1.17\pi$ \\
\noalign{\smallskip}\toprule
\end{tabular}
\end{table}}\label{table1}

\begin{figure}[tbp]
\centering
\includegraphics[width=1.0\linewidth]{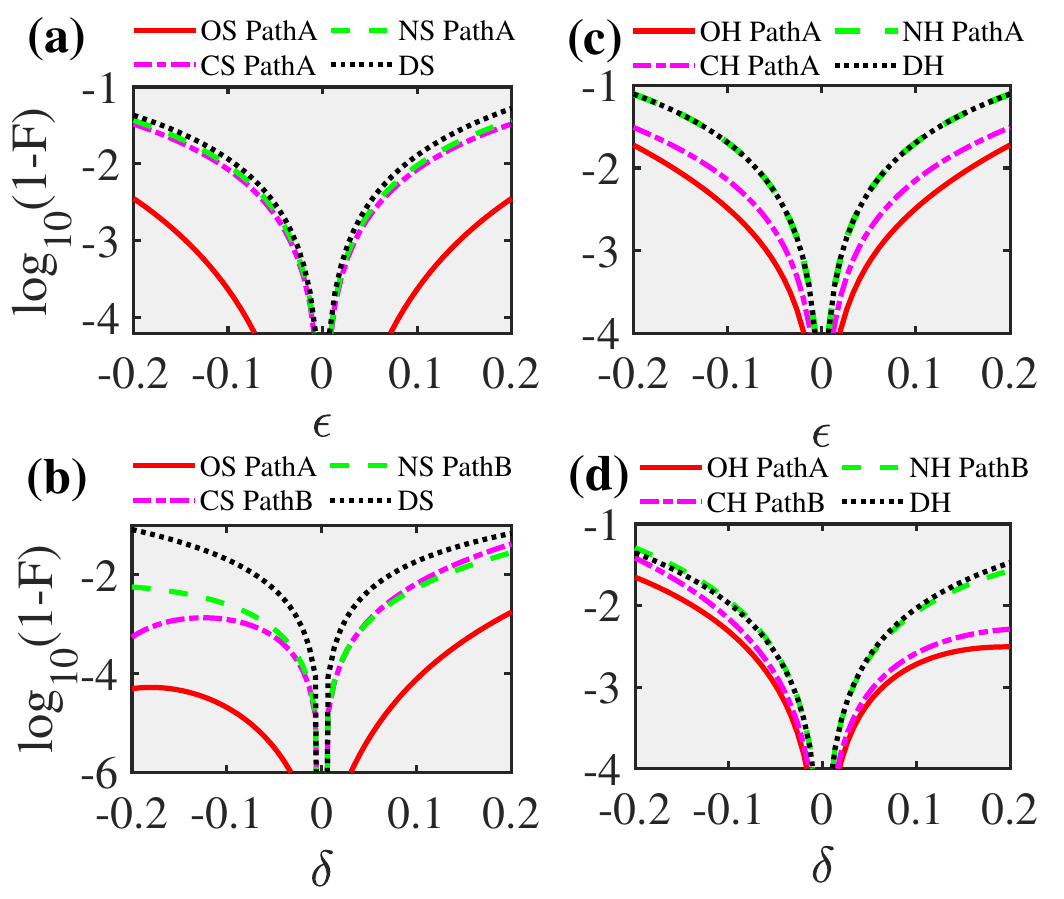}
\caption{The robustness comparison of our optimized composite nonadiabatic geometric S (OS) and H (OH) gates within Path A against [(a),(c)] X and [(b),(d)] Z errors, respectively, on the optimal parameters with conventional composite nonadiabatic geometric S (CS) and H (CH) gates, nonadiabatic geometric S (NS) and H (NH) gates, and corresponding dynamical S (DS) and H (DH) gates, where $\epsilon,\delta\in[-0.2,0.2]$ and decoherence is not considered.}\label{Figure3}
\end{figure}

\subsection{The gate robustness}
In the subsection, based on these optimal parameters, we will compare the robustness of our OCNGQGs against certain errors with those of conventional CNGQGs \cite{chen2018nonadiabatic,zhou2021nonadiabatic}, NGQGs \cite{zhao2017rydberg,xu2020experimental} and dynamical gates (DGs) \cite{PhysRevLett.116.020501}. Notably, NGQGs within the orange-slice-shaped evolution path have two configurations, ``Path A" and ``Path B", which can resist X and Z errors, respectively. CNGQGs also exhibit two configurations by repeating the original path twice, enhancing robustness against individual errors. The constructions for geometric and dynamical gates are detailed in Appendix B and Appendix C, respectively. Our OCNGQGs follow a similar approach but can resist both X and Z errors at their optimal parameters $k$. For demonstration, we use ``Path A" for OCNGQGs in the main text, while ``Path B" is described in Appendix A.

As shown in Fig. 3, we have plotted the curves depicting how the infidelity of S and H quantum gates changes with error rates $\epsilon$ or $\delta$, based on three geometric schemes and the dynamical scheme. Specifically, Figs. 3(a) and 3(c) illustrate that our optimized composite geometric scheme for S and H gates with ``Path A" are more robust against X error compared to CNGQGs and NGQGs within ``Path A", and the corresponding DGs. Conversely, for Z errors, Figs. 3(b) and 3(d) demonstrate that our scheme also offers significant advantages over CNGQGs and NGQGs within ``Path B", and corresponding DGs.

\begin{figure}[tbp]
\centering
\includegraphics[width=1.0\linewidth]{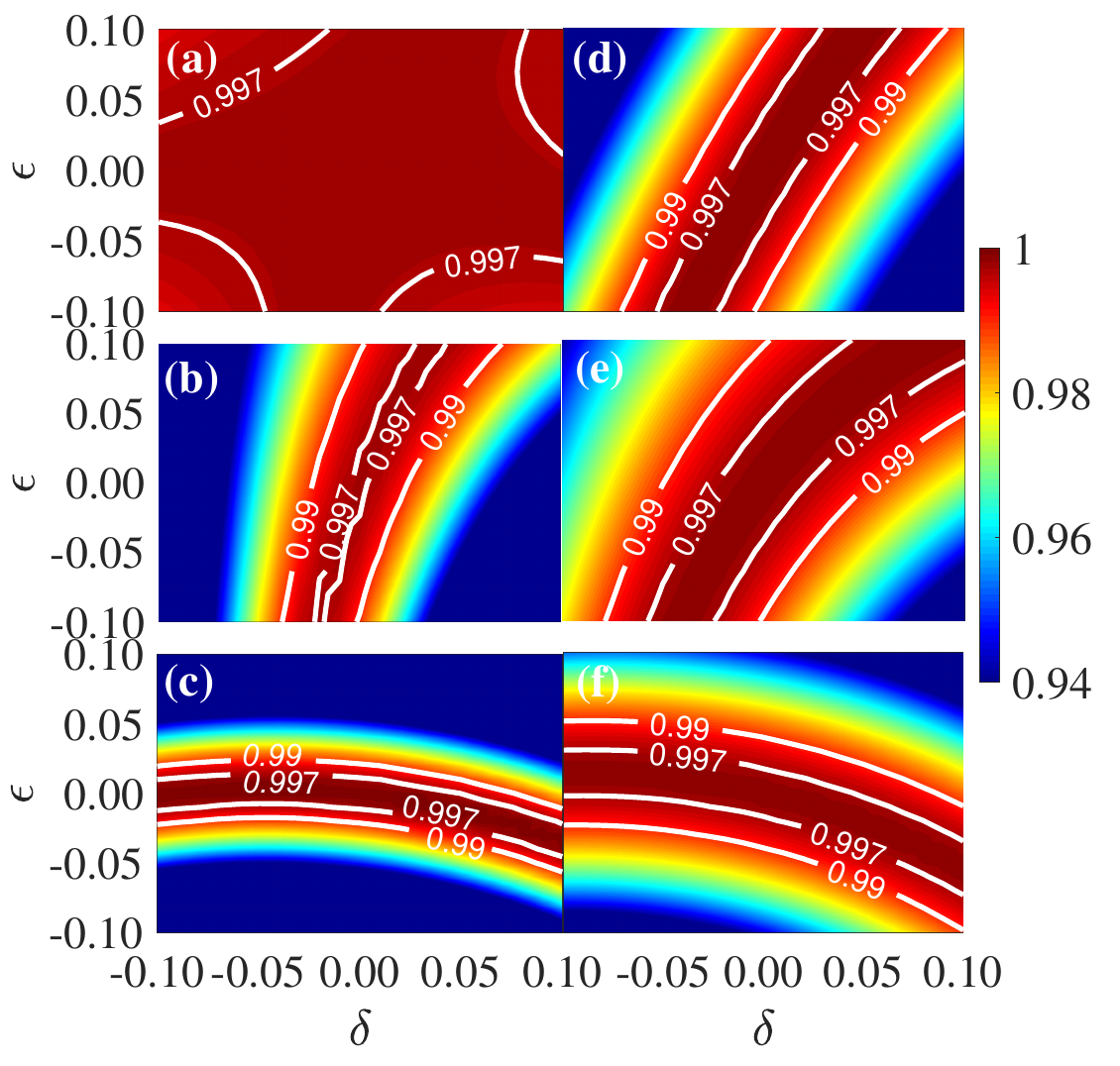}
\caption{The performance of S gate with X and Z errors for six schemes, including (a) OCNGQG, (b),(e) CNGQG and NGQG within Path A, (c),(f) CNGQG and NGQG within Path B, and (d) DG schemes, in which the error rates $\epsilon,\delta\in[-0.1,0.1]$ and decoherence rates $\kappa_z=\kappa_-=\Omega_m/10^4$. }\label{Figure4}
\end{figure}

Given that the gate times of our OCNGQGs are at least twice as long as those of conventional NGQGs, the decoherence effect cannot be ignored. Consider the decoherence rates $\kappa_z=\kappa_-=\Omega_m/10^4$ and the pulse shape $\Omega(t)=\Omega_m \sin^2(\pi t/\tau)$, in Figs. 4 and 5, we present the gate-fidelity variations of S and H gates under both X and Z noises, respectively, based on six different schemes including our OCNGQG, CNGQG and NGQG within ``Path A" \cite{chen2018nonadiabatic}, CNGQG and NGQG within ``Path B" \cite{zhou2021nonadiabatic}, and DG \cite{PhysRevLett.116.020501} schemes. By comparing all the subgraphs in Fig. 4, it is evident that our geometric S gate is more robust than others in resisting X and Z errors simultaneously under decoherence. However, we can find, by comparing Figs. 4(b) and 4(e), the composite geometric scheme based on ``Path A" can effectively enhance the robustness against X error but weaken the one against Z error. This is consistent with the study in Ref. \cite{chen2018nonadiabatic}. Similarly, the composite geometric scheme based on ``Path B" can enhance the robustness against Z error but weaken that for X error \cite{zhou2021nonadiabatic}, as seen in Figs. 4(c) and 4(f). The results in Fig. 5 demonstrate a similar conclusion that the robustness of our scheme for the H gate can outperform other schemes overall. But, the robustness advantage is not obvious compared with the S gate in Fig. 4. The reason is that the optimal $k$ value for resisting X and Z errors is not consistent for the H gate, as shown in Table I, resulting in the optimal $k$ value not being truly optimal for X or Z errors when considering both types of errors simultaneously. In addition, due to the longer evolution time of the H gate comparing with the S gate, they are also susceptible to more decoherence effects, leading to a decrease in gate fidelity. Furthermore, the optimal path parameters $k$ used here are still the same as in Table I, as different choices of $k$ values do not change the entire length of the evolution path. That is, $k$ only relates to specific systematic errors instead of specific system parameters.

\begin{figure}[tbp]
\centering
\includegraphics[width=1.0\linewidth]{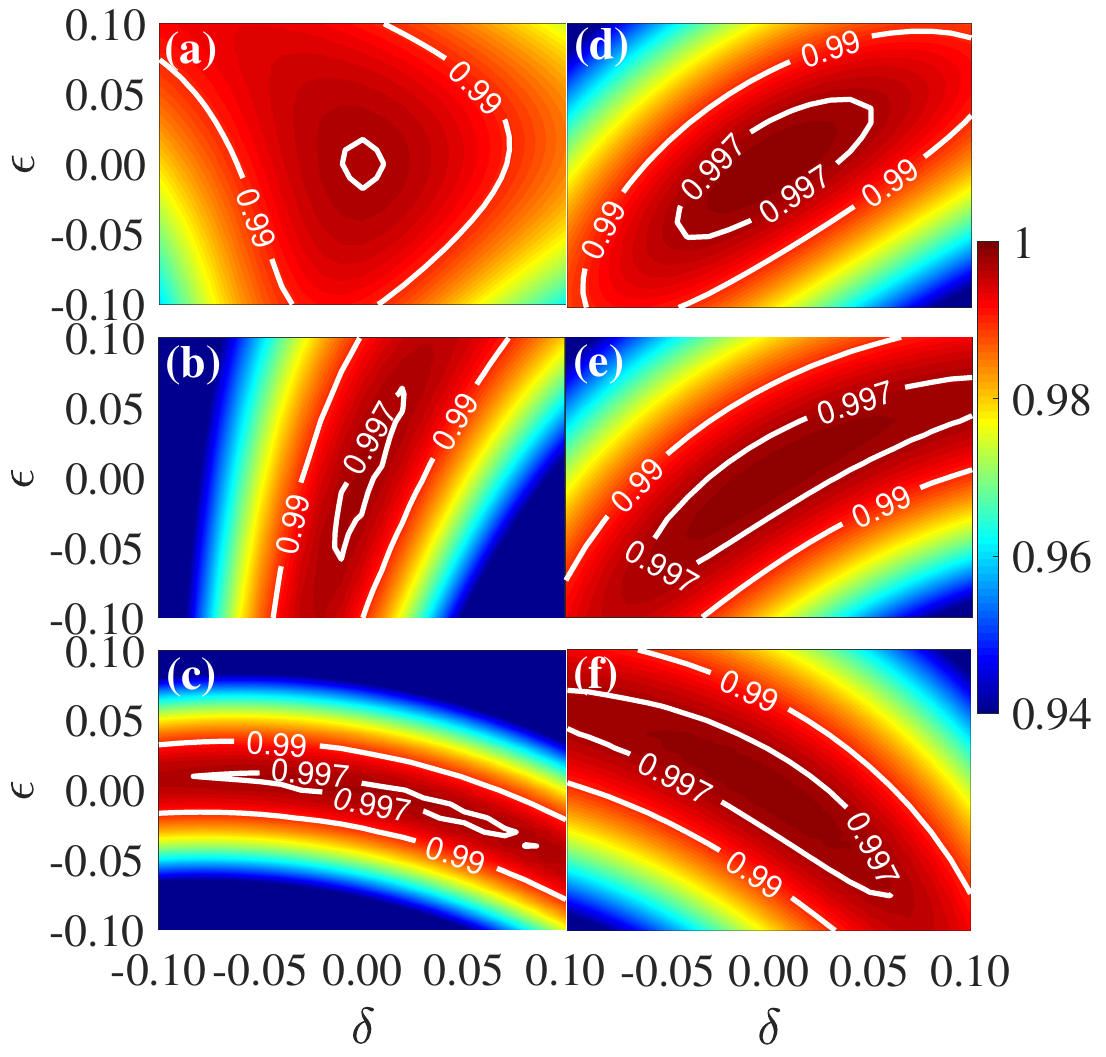}
\caption{The performance of H gate with X and Z errors for six schemes, including (a) OCNGQG, (b),(e) CNGQG and NGQG within Path A, (c),(f) CNGQG and NGQG within Path B, and (d) DG schemes, in which the error rates $\epsilon,\delta\in[-0.1,0.1]$ and decoherence rates $\kappa_z=\kappa_-=\Omega_m/10^4$. }\label{Figure5}
\end{figure}

\section{Physical implementation}
In this section, we propose to implement the above scheme on superconducting quantum circuits, which includes the cases for both single-qubit gates and a nontrivial two-qubit gate with current superconducting quantum control techniques.

\subsection{Single-qubit gates}
Consider a superconducting transmon qubit, the energy spectrum of which is shown in Fig. 6(a). It is driven by a microwave field, and the full Hamiltonian can be given by
\begin{eqnarray}
\begin{split}
\mathcal{H}_0(t)&=\sum_{l=1}^{+\infty}\Big\{l\omega-\frac{l(l-1)}{2}\alpha|l\rangle\langle l| \\
&+\big[\frac{\sqrt{l}}{2}\Omega_R(t)|l-1\rangle\langle l|e^{i(\omega_d t-\phi)}+\textmd{H.c.}\big] \Big\},
\end{split}
\end{eqnarray}
where $\omega$ and $\alpha$ are the qubit frequency and anharmonicity of the transmon, $\Omega_R(t)$, $\omega_d$ and $\phi$ are the amplitude, frequency and phase of the microwave field, respectively.  As transmons have weak anharmonicity, we could suppress the information leakage from the computational subspace  by utilizing derivative removal via adiabatic gate (DRAG) technology \cite{PhysRevLett.103.110501,PhysRevA.83.012308}. That is, the $\Omega_R(t)$ is set as $\Omega_R(t)=\Omega(t)-i\dot{\Omega}(t)/2\alpha$ with $\Omega(t)=\Omega_m\sin^2(\pi t/\tau)$ being the experiment-friendly original pulse. Additionally, the original Hamiltonian $\mathcal{H}_0(t)$ is rotated into a frame with respect to the driving frequency $\omega_d$ with $\mathcal{R}=e^{-i\omega_d t\sum_{l=1}^{+\infty}l|l\rangle\langle l|}$, where a resonant condition $\omega=\omega_d$ is assumed. Then, the reduced Hamiltonian in the computational subspace $\{|0\rangle,|1\rangle\}$ takes the same form as Eq. (1). Thus, by setting the same parameter constraints as in Eq. (6), the OCNGQGs can be obtained.

\begin{figure}[tbp]
\centering
\includegraphics[width=1.0\linewidth]{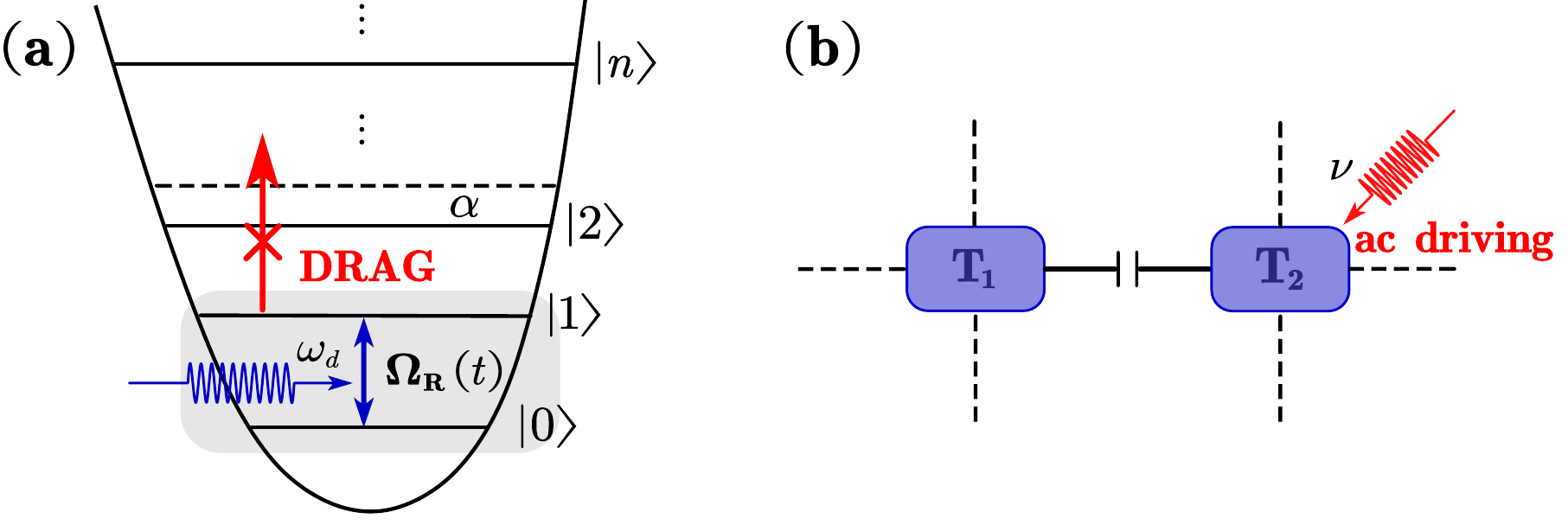}
\caption{Illustration of our implementation. (a) The energy spectrum of a resonantly driven transmon qubit, with a weak anharmonicity $\alpha$. The leakage errors for higher energy levels can be suppressed by DRAG technology. (b) Schematic diagram of capacitive coupling between two transmons, where ac microwave driving is applied to $T_2$ to achieve controllable coupling between them. }\label{Figure6}
\end{figure}

Similarly, we numerically simulate the gate performance by the Lindblad master equation, which takes the same form as in Eq. (9), except for changing $\mathcal{H}'_1(t)$ to $\mathcal{H}_0(t)$, in order to fully describe the dynamical process of the transmon. Here, $\mathcal{A}_z=\sum_{l=1}^{+\infty}l|l\rangle\langle l|$ and $\mathcal{A}_-=\sum_{l=1}^{+\infty}\sqrt{l}|l-1\rangle\langle l|$, and $\Omega_m=2\pi\times60$ MHz, $\alpha=2\pi\times320$ MHz and the decoherence rates $\kappa_z=\kappa_-=2\pi\times2$ kHz are within the current available parameter range \cite{kjaergaard2020superconducting,wang2022towards}. We take the S gate as an example and test its robustness for different schemes, and the X and Z errors are set as $\Omega(t)\rightarrow(1+\epsilon)\Omega(t)$ and $\delta\Omega_m(|1\rangle\langle 1|+2|2\rangle\langle 2|)$, respectively. As shown in Fig. 7, at the optimal parameter $k\simeq1.13\pi$, the robustness of S gate based on our scheme is significantly better than the other five schemes, resisting X and Z errors simultaneously under decoherence. Furthermore, since the robustness of the H gate with transmon is almost the same as that in Fig. 5, it is not shown here. The performance comparison of these schemes is listed in Table II. It can be found that our scheme possesses much stronger gate robustness against various errors under decoherence, despite the longer gate duration. Moreover, it can be predicted that with the further increase of coherence time, the robustness of our scheme will be stronger.

In addition, in our simulations, the pulse shape can be arbitrary under a device-limited peak, and thus we can set them in an experimentally friendly way. Meanwhile, we also utilize the DRAG technology to suppress the leakage errors from non-computational subspaces for high gate fidelity. So, our scheme is feasible in realistic superconducting systems. Besides, our scheme is also compatible with more complex pulse waveforms. That is, it can be further combined with pulse-shaping techniques to improve gate fidelity and robustness.

\setlength{\tabcolsep}{7.8mm}{\begin{table*}
\centering
\caption{The gate performance comparison for various schemes in superconducting transmon systems, where the performance indicators include gate duration, fidelity under decoherence, and robustness under both given errors and decoherence. The system parameters of all schemes are set to be the same as in Sec. III, including decoherence $\kappa=2\pi\times2$ kHz.}
\begin{tabular}{ccccc}
\hline\hline\noalign{\smallskip}
\multicolumn{2}{c}{  Schemes$\diagdown$ Performances} & Gate duration& Fidelity &Robustness \\
\noalign{\smallskip}\hline\noalign{\smallskip}
\multirow{2}{*}{OCNGQGs}&$\textrm{S}$ & $66.7$ns & $0.9996$ & $\epsilon=0.1$, $0.9994$;$\delta=0.1$, $0.9991$\\
 &$\textrm{CP}$ & $121.6$ns & $0.9970$& $\epsilon=0.2$, $0.9928$;$\delta=0.2$, $0.9968$ \\
\noalign{\smallskip}\hline\noalign{\smallskip}
\multirow{3}{*}{CNGQGs}&$\textrm{S for Path A}$ & $66.7$ns & $0.9996$ & $\epsilon=0.1$, $0.9912$\\
 &$\textrm{S for Path B}$ & $66.7$ns & $0.9994$& $\delta=0.1$, $0.9689$ \\
 &$\textrm{CP}$ & $121.6$ns & $0.9972$& $\epsilon=0.2$, $0.9661$;$\delta=0.2$, $0.9743$ \\
  \noalign{\smallskip}\hline\noalign{\smallskip}
\multirow{3}{*}{NGQGs}&$\textrm{S for Path A}$ & $33.3$ns & $0.9998$ & $\epsilon=0.1$, $0.9906$\\
 &$\textrm{S for Path B}$ & $33.3$ns & $0.9997$& $\delta=0.1$, $0.9835$ \\
 &$\textrm{CP}$ & $60.8$ns & $0.9987$& $\epsilon=0.2$, $0.9500$;$\delta=0.2$, $0.9937$ \\
   \noalign{\smallskip}\hline\noalign{\smallskip}
\multirow{2}{*}{DGs}&$\textrm{S}$ & $27.3$ns & $0.9998$ & $\epsilon=0.1$, $0.9878$;$\delta=0.1$, $0.9447$\\
 &$\textrm{CP}$ & $60.8$ns & $0.9980$& $\epsilon=0.2$, $0.9652$;$\delta=0.2$, $0.9926$ \\
\noalign{\smallskip}\toprule
\end{tabular}
\end{table*}}\label{table2}

\begin{figure}[tbp]
\centering
\includegraphics[width=1.0\linewidth]{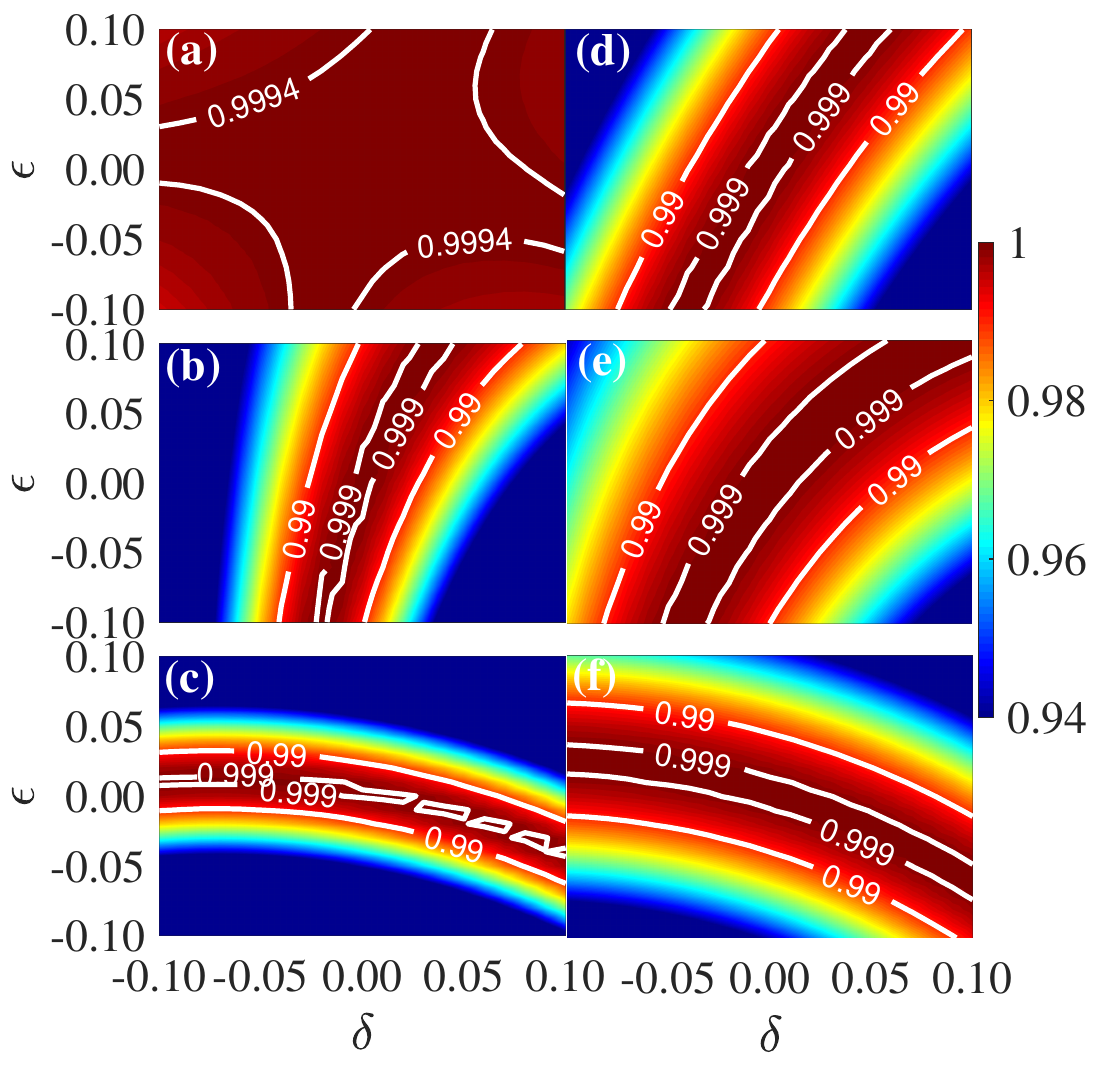}
\caption{The robustness of S gate against both X and Z errors on a transmon qubit with decoherence rates $\kappa_z=\kappa_-=2\pi\times2$ kHz for six schemes, including (a) OCNGQG, (b),(e) CNGQG and NGQG within Path A, (c),(f) CNGQG and NGQG within Path B, and (d) DG schemes, in which the error rates $\epsilon,\delta\in[-0.1,0.1]$.}\label{Figure7}
\end{figure}

\begin{figure}[tbp]
\centering
\includegraphics[width=1.01\linewidth]{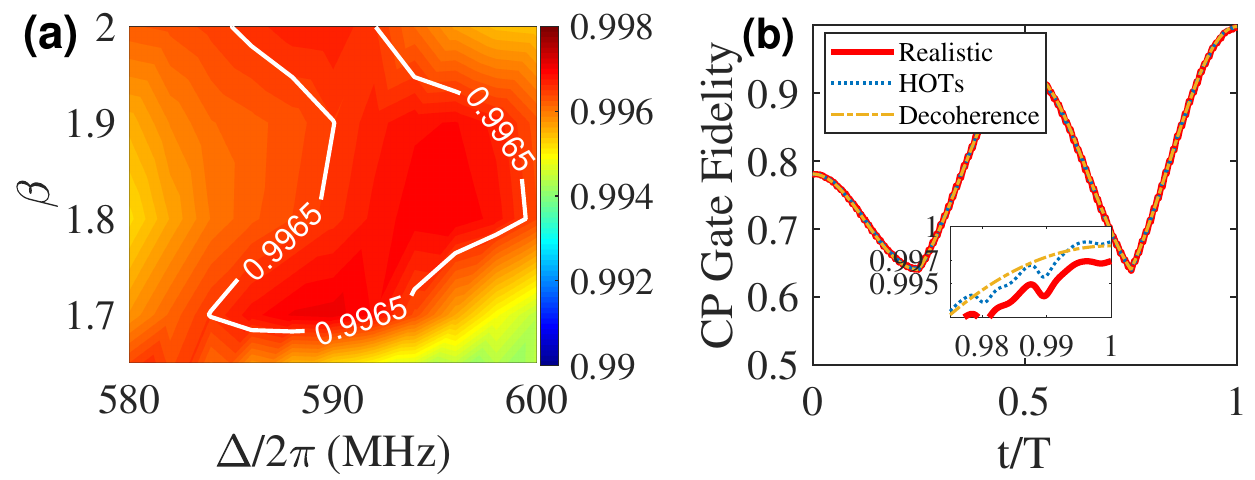}
\caption{The fidelity of implemented two-qubit CP gate with $\gamma'_g=\pi/2$. (a) Within the available settings of experimental parameters, the gate fidelity of CP gate alters with parameters $\Delta$ and $\beta$ under optimal path parameter $k\simeq1.27\pi$. (b) The realistic dynamics of gate fidelity for CP gate with optimal parameters of (a), where we also consider the impact of high-order oscillating terms (HOTs) errors and decoherence effect on the fidelity, respectively.}\label{Figure8}
\end{figure}

\subsection{A nontrivial two-qubit gate}
To achieve universal quantum computing, in addition to a set of universal single-qubit quantum gates, a nontrivial two-qubit quantum gate is also required. Here, we construct an optimized composite controlled-phase (CP) gate based on two capacitive coupled transmons, as shown in Fig. 6(b). The system Hamiltonian can be expressed as
\begin{eqnarray}
\begin{split}
\mathcal{H}_{12}(t)&=\sum_{j=1}^{2}\sum_{l=1}^{+\infty}\big[\omega_jl-\frac{l(l-1)}{2}\alpha_j|l\rangle_j\langle l|\big]  \\
&\quad+g_{12}(\sigma_1\sigma_2^{\dagger}+\textmd{H.c.} )
\end{split}
\end{eqnarray}
in which $\sigma_j=\sum_{l=0}^{+\infty}\sqrt{l+1}|l\rangle_j\langle l+1|$ is the lowering operator for the $(l+1)$-th energy level of transmon $T_j$; $\omega_j$, $\alpha_j$ denote the transition frequency and inherent anharmonicity of $T_j$, respectively; $g_{12}$ is the fixed interaction strength between $T_1$ and $T_2$ when their transition frequencies are constant. To achieve parametric controllable coupling \cite{PhysRevA.96.062323,reagor2018demonstration,PhysRevApplied.13.064012}, we apply an ac microwave driving on $T_2$ to modify its transition frequency to be time-dependent, i.e., $\omega_2(t)=\omega_2+\beta\nu\sin(\nu t+\eta)$, where $\nu$, $\eta$, and  $\beta\nu$ are the frequency, phase and amplitude of the driving field, respectively, with $\beta$ being an dimensionless parameter. By this way, in the interaction picture, the interaction Hamiltonian between the two coupled transmons is
\begin{eqnarray}
\begin{split}
\mathcal{H}'_{12}(t)&=g_{12}\Big\{\big[|10\rangle_{12}\langle01|e^{i\Delta t}+\sqrt{2}|20\rangle_{12}\langle11|e^{i(\Delta-\alpha_1)t}        \\ &+\sqrt{2}|11\rangle_{12}\langle02|e^{i(\Delta+\alpha_2)t}\big]e^{i\beta\cos(\nu t+\eta)}+\textmd{H.c.}     \Big\},
\end{split}
\end{eqnarray}
where $\Delta=\omega_1-\omega_2$ is the detuning between transition frequencies of $T_1$ and $T_2$, respectively. To derive the effective Hamiltonian in the two-qubit subspace $\{|11\rangle,|02\rangle\}$, we approximate Eq. (13) using the Jacobi-Anger identity of $\exp{[i\beta\cos{(\nu t+\eta)}]}=\sum_{n=-\infty}^{+\infty}i^nJ_n(\beta)\exp[i n(\nu t+\eta)]$, with $J_n(\beta)$ being Bessel functions of the first kind. When adjusting the driving frequency $\nu$ to satisfy $\nu=\Delta+\alpha_2$ and neglecting the high-frequency oscillation terms, the effective Hamiltonian can be calculated as
\begin{equation}
\mathcal{H}_{\textrm{eff}}(t)=\frac{g_{\textrm{eff}}}{2}\left[\cos(\eta-\frac{\pi}{2})\tilde{\sigma}_x+\sin(\eta-\frac{\pi}{2})\tilde{\sigma}_y\right],
\end{equation}
where the effective coupling $g_\textrm{eff}=2\sqrt{2}g_{12}J_1(\beta)$ can be adjusted by the parameter $\beta$, and $\tilde{\sigma}_x$ and $\tilde{\sigma}_y$ are Pauli matrices in subspace $\{|11\rangle,|02\rangle\}$. It is evident that the form of $\mathcal{H}_{\textrm{eff}}(t)$ matches that of $\mathcal{H}_1(t)$ in Sec. II A. Therefore, we can implement the optimized composite geometric two-qubit CP gates. Specifically, the state $|11\rangle$ can accumulate a pure geometric phase $-\gamma'_g$ via the application of optimized composite geometric pulses. The implemented CP gates, in the two-qubit computational subspace $\{|00\rangle,|01\rangle,|10\rangle,|11\rangle\}$, is
\begin{equation}
U(\tau')=\textrm{diag}(1,1,1,e^{-i\gamma'_g}).
\end{equation}

To fully demonstrate the performance of the geometric CP gates, we use $\gamma'_g=\pi/2$ as an example, and numerically simulate it using a two-qubit Lindblad master equation of
\begin{equation}
\dot{\rho}_2(t)=i[\rho_2(t),\mathcal{H}'_{12}(t)]+\frac{1}{2}\sum_{j=1}^{2}\sum_{m=z,-}\kappa_m^j\mathcal{L}(\mathcal{A}_m^j),
\end{equation}
where $\mathcal{A}_-^j=\sum_{l=1}^{+\infty}\sqrt{l}|l-1\rangle_j\langle l|$ and $\mathcal{A}_z^j=\sum_{l=1}^{+\infty}l|l\rangle_j\langle l|$, and $\kappa_z^j$ and $\kappa_-^j$ are the dephasing and decay rates of the transmon $T_j$, respectively. Besides, the gate fidelity for two-qubit case can be also defined as $F'=(1/16)\sum_{m=1}^{16}\langle\psi'_m(0)|U^{\dagger}(\tau')\rho_2(\tau')U(\tau')|\psi'_m(0)\rangle$ with $\rho_2(t)$ being the density matrix for two qubits. Furthermore, $|\psi'_m(0)\rangle=|\psi'_{m1}(0)\rangle\otimes|\psi'_{m2}(0)\rangle$ is set as the initial state, with $|\psi'_{mj}(0)\rangle$ ($j=1,2$) belonging to single-qubit set of $\{|0\rangle,|1\rangle,(|0\rangle-i|1\rangle)/\sqrt{2},(|0\rangle+|1\rangle)/\sqrt{2}\}$.

\begin{figure}[tbp]
\centering
\includegraphics[width=1.01\linewidth]{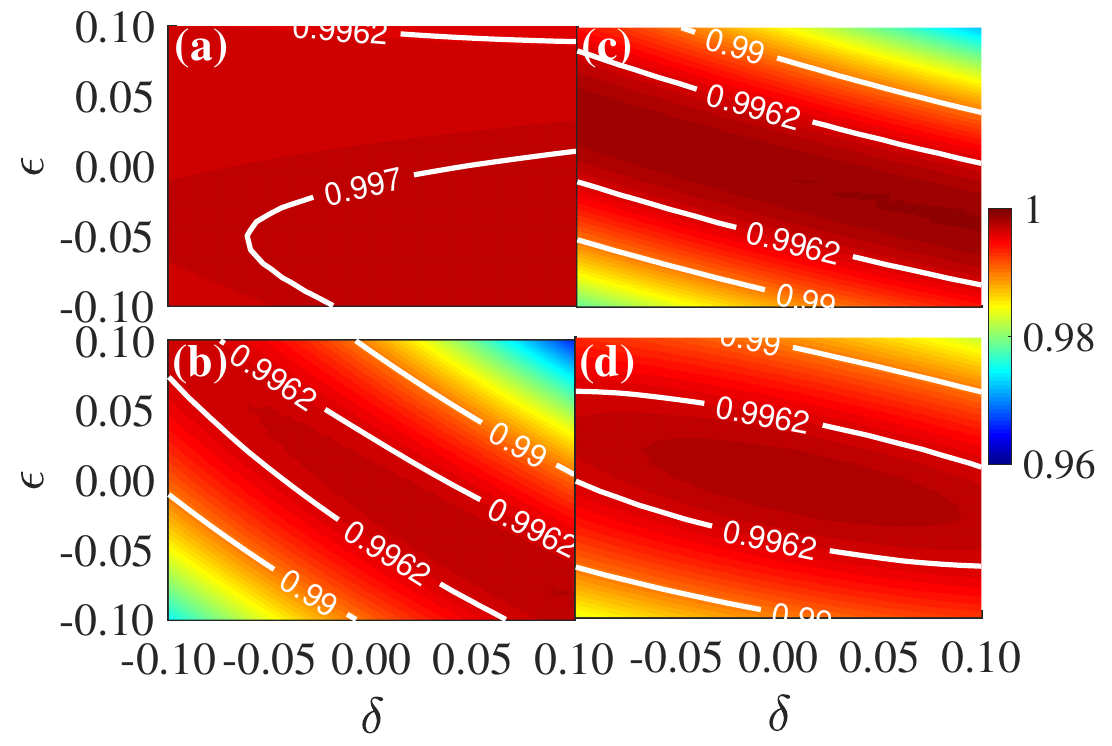}
\caption{The robustness of two-qubit CP gate against both X and Z errors in superconducting transmon systems with decoherence rates $\kappa^i_z=\kappa^i_-=2\pi\times2$ kHz $(i=1,2)$ for (a) OCNGQG, (b),(c) CNGQG and NGQG within Path A, and (d) DG, respectively, where the error rates $\epsilon,\delta\in[-0.1,0.1]$. }\label{Figure9}
\end{figure}

Under the optimal path parameter $k\simeq1.27\pi$, we set reasonable experimental parameters $g_{12}=2\pi\times10$ MHz, $\alpha_1=2\pi\times320$ MHz, $\alpha_2=2\pi\times300$ MHz and $\kappa_z^1=\kappa_z^2=\kappa_-^1=\kappa_-^2=2\pi\times2$ kHz \cite{kjaergaard2020superconducting}, and test the gate fidelity as a function of parameters $\Delta$ and $\beta$ in Fig. 8(a). We find that the optimal parameters are around $\Delta=2\pi\times594$ \textrm{MHz} and $\beta=1.8$. The corresponding CP gate fidelity can reach $99.70\%$, whose dynamics is shown in Fig. 8(b). Moreover, in the Fig. 8(b), we also plot the impact of high-order oscillating terms errors and decoherence effect on the fidelity of the two-qubit gate, respectively. It can be observed that the infidelity arises from both approximately attributing $0.13\%$ and $0.17\%$, respectively. In addition, as shown in Fig. \ref{Figure9}(a), we also test the robustness of the CP gate against both X and Z errors, and compare it with other schemes in Figs. 9(b)-9(d).  Fig. \ref{Figure9} shows that our scheme is significantly superior to other three ones. Note that, the X and Z errors for two-qubit case are considered as $g_{12}\rightarrow(1+\epsilon)g_{12}$ and $\Delta\rightarrow\Delta+\delta g_{12}$, respectively. In Table II, we give the CP gate performance indicators of different schemes. Through comparison, we find that our scheme has a longer gate time than schemes of NGQGs and DGs, but the gate robustness is the best among all schemes.

In addition, for $\gamma'_g=\pi/2$, due to the fact that conventional composite geometric CP gate within ``Path B" has the same evolution path for one within ``Path A", while geometric CP within ``Path B" and the one within ``Path A" differ by a negative sign on state $|11\rangle$, which results in different quantum gates, only the path configuration A is shown in Fig. \ref{Figure9}.

\begin{figure*}[tbp]
\centering
\includegraphics[width=0.9\linewidth]{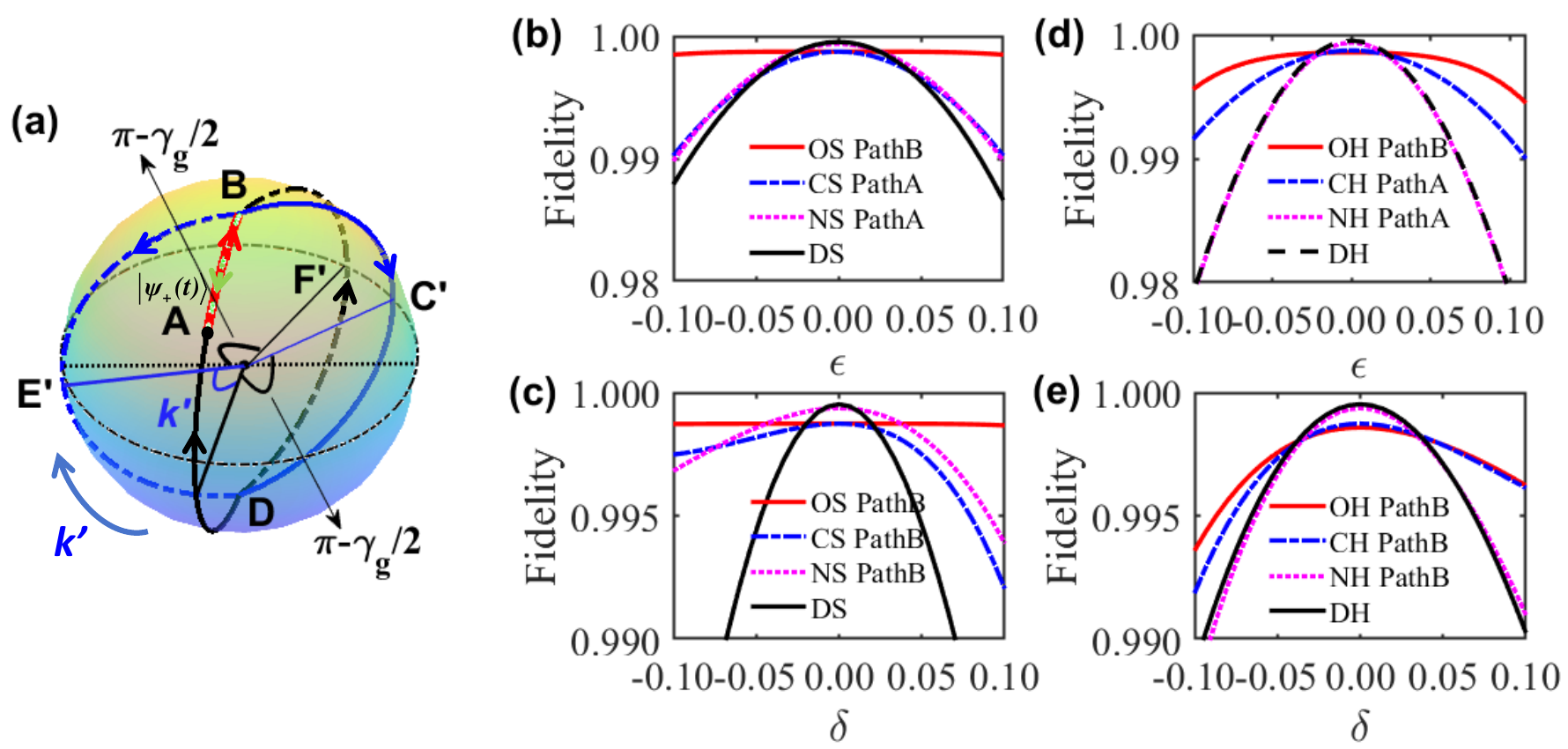}
\caption{OCNGQGs for Path B. (a) Schematic representation of closed evolution trajectory for Path B, where the loop follows $\textrm{AB}-\textrm{BC$'$ D}-\textrm{DA}-\textrm{AB}-\textrm{BE$'$D}-\textrm{DF$'$B}-\textrm{BA}$ and $k'$ is an optimizable path parameter. Under the decoherence rates $\kappa_z=\kappa_-=\Omega_m/10^4$, the robustness comparison of our optimized composite nonadiabatic geometric S (OS) and H (OH) gates within Path B against [(b),(d)] X and [(c),(e)] Z errors, respectively, on the optimal parameters with conventional composite nonadiabatic geometric S (CS) and H (CH) gates, nonadiabatic geometric S (NS) and H (NH) gates, and corresponding dynamical S (DS) and H (DH) gates, where $\epsilon,\delta\in[-0.1,0.1]$.  }\label{Figure10}
\end{figure*}

\section{Conclusion}
In summary, we have developed a novel scheme for OCNGQGs using the superconducting transmon systems. This approach extends conventional composite geometric schemes and addresses their limitation of suppressing error in only one direction. By optimizing the path parameter $k$, our solution can resist noise sources in specific directions and simultaneously in two directions. Numerical simulations show that our scheme outperforms conventional composite and single-loop geometric schemes, as well as corresponding dynamical ones, in resisting X and Z errors simultaneously. Additionally, our optimization scheme can be naturally extended to the two-qubit scenario, with similar simulation results verified. Thus, our protocol offers a new optimization method for conventional composite geometric scheme, making it more promising to achieve robust quantum gates for future scalable quantum computation.

\begin{acknowledgements}
This work is supported by the National Natural Science Foundation of China (Grant No. 12275090), the key Scientific Research Foundation of Anhui Provincial Education Department (Grant No.2023AH050481), the Guangdong Provincial Quantum Science Strategic Initiative (Grant No. GDZX2203001), and the Research Foundation for Advanced Talents of WAU (Grant No. WGKQ2021004).

\end{acknowledgements}

\appendix
\section{The OCNGQG scheme for Path B}
In addition to enabling OCNGQCs under the Path A, our solution can also be implemented through Path B. Specifically, as illustrated in Fig. \ref{Figure10}(a), the  evolution trajectory is also divided into seven parts, and follows the loop of $\textrm{AB}-\textrm{BC$'$ D}-\textrm{DA}-\textrm{AB}-\textrm{BE$'$D}-\textrm{DF$'$B}-\textrm{BA}$. Unlike Path A, the inner angles for two orange-petal paths are both $\pi-\gamma_g/2$ instead of $\gamma_g/2$, and the angle between them is the optimized  parameter $k'$. Based on Eq. (3), the Hamiltonian parameters need to be set as
\begin{equation}
\left\{
     \begin{array}{lr}
     \phi_2\longrightarrow \phi'_2=\beta_0-\frac{\gamma_g}{2}-\frac{\pi}{2}, &  \\
     \phi_5\longrightarrow \phi'_5=\beta_0-k'+\frac{\pi}{2}, & \\
     \phi_6\longrightarrow \phi'_6=\beta_0-k'+\frac{\gamma_g}{2}+\frac{\pi}{2}, & \\
     \end{array}
\right.
\end{equation}
with other parameters being the same as Eq. (6). The obtained evolution operator is $e^{-i\gamma_g\vec{n}\cdot\vec{\sigma}}$, coincides with that of Path A.

Similarly, we can numerically determine the optimal path parameter $k'$ for X and Z errors. 
In Figs. 10(b)-10(e), we compare the robustness of our S and H gates with that of three other schemes under X and Z errors, respectively, where the rates of decoherence and errors are set as $\kappa_z=\kappa_-=\Omega_m/10^4$ and $\epsilon,\delta\in[-0.1,0.1]$. It can be observed from Figs. 10(b) and 10(c) that our optimized geometric S gate, with $k'\simeq1.87\pi$, exhibits significantly enhanced robustness compared to other schemes in terms of X or Z errors, despite having a longer gate time than the corresponding conventional NGQG and DG. For our H gate, it maintains a clear advantage over other schemes in resisting X errors while retaining an overall advantage in resisting Z errors, as illustrated in Figs. 10(d) and 10(e). Notably, the optimal $k'$ values at this point can be searched for separately, yielding $1.67\pi$ and $0.27\pi$, respectively.

\section{The construction of NGQGs and CNGQGs}
Conventional NGQGs are based on the orange-slice-shaped single-loop evolution path \cite{zhao2017rydberg,chen2018nonadiabatic}, which are divided into three segments. Note that the form of system Hamiltonian is the same as that in Eq. (1), then the corresponding parameters of driving microwave pulses need to meet the requirements of
\begin{eqnarray}
\begin{split}
\int_0^{\tau_1}\Omega(t)dt&=\alpha_0, \phi_1=\beta_0-\frac{\pi}{2}, t\in[0,\tau_1],\\
\int_{\tau_1}^{\tau_2}\Omega(t)dt&=\pi, \phi_2=\beta_0-\gamma_g+\frac{\pi}{2},t\in[\tau_1,\tau_2],\\
\int_{\tau_2}^{\tau}\Omega(t)dt&=\pi-\alpha_0, \phi_3=\beta_0-\frac{\pi}{2}, t\in[\tau_2,\tau].\\
\end{split}
\end{eqnarray}
Hence, the evolution operator at the final time $\tau$ is
\begin{eqnarray}
U(\tau)=U(\tau,\tau_2)U(\tau_2,\tau_1)U(\tau_1,0)  
=e^{-i\gamma_g\vec{n}\cdot\vec{\sigma}},
\end{eqnarray}
which is aligns with Eq. (7) in the main text, indicating that any single-qubit quantum gate can be implemented. Furthermore, the evolution loop consists solely of meridians, leading to $\dot{\beta}(t)=0$. Consequently, the dynamical phase is consistently zero, and $U(\tau)$ is a pure geometric quantum gate. We call it as NGQGs within Path A, whose inner angle of orange-slice-shaped trajectory is the geometric phase $\gamma_g$. Contrarily, if the inter angle is $\pi-\gamma_g$, we call it as NGQGs within Path B, necessitating an adjustment of the pulse phase of the second segment to $\beta_0-\gamma_g-\pi/2$. Then, the corresponding evolution operator is $U'(\tau)=-e^{-i\gamma_g\vec{n}\cdot\vec{\sigma}}$, equals to $U(\tau)$ in Path A.

For conventional CNGQGs with two distinct path configurations, we adopt NQGQs with Path A $U_1(\gamma_c)$ or Path B $U_2(\gamma_c)$ as building blocks, iteratively cascading them $n$ times. This iteration corresponds to the repetition of the evolution loop $n$ times. By this way, the CNGQGs can be obtained as
\begin{equation}
U_c^x\left(\gamma_g\right)=\left[U_x(\gamma_c)\right]^n=\left[(-1)^{x+1}e^{-i\gamma_c\vec{n}\cdot\vec{\sigma}}\right]^n,
\end{equation}
where $x=1,2$ corresponds CNGQGs within Path A and Path B, respectively, and $\gamma_c=\gamma_g/n$. Specifically, we take two-loop CNGQGs as an example, which is adopted in the main text. The Hamiltonian remains identical to $\mathcal{H}_1(t)$, and the required parameter settings for Path A are
\begin{eqnarray}
\int_0^{\tau_1}\Omega(t)dt&=&\alpha_0, \phi_1=\beta_0-\frac{\pi}{2}, t\in[0,\tau_1],\notag\\
\int_{\tau_1}^{\tau_2}\Omega(t)dt&=&\pi, \phi_2=\beta_0-\frac{\gamma_g}{2}+\frac{\pi}{2},t\in[\tau_1,\tau_2],\notag\\
\int_{\tau_2}^{\tau_3}\Omega(t)dt&=&\pi-\alpha_0, \phi_3=\beta_0-\frac{\pi}{2}, t\in[\tau_2,\tau_3].\notag\\
\int_{\tau_3}^{\tau_4}\Omega(t)dt&=&\alpha_0, \phi_4=\beta_0-\frac{\pi}{2}, t\in[\tau_3,\tau_4],\notag\\
\int_{\tau_4}^{\tau_5}\Omega(t)dt&=&\pi, \phi_5=\beta_0-\frac{\gamma_g}{2}+\frac{\pi}{2},t\in[\tau_4,\tau_5],\notag\\
\int_{\tau_5}^{\tau}\Omega(t)dt&=&\pi-\alpha_0, \phi_6=\beta_0-\frac{\pi}{2}, t\in[\tau_5,\tau].
\end{eqnarray}
As a result, the evolution operator can be calculated as
\begin{equation}
U^1_c(\tau)=e^{-i\frac{\gamma_g}{2}\vec{n}\cdot\vec{\sigma}}\cdot e^{-i\frac{\gamma_g}{2}\vec{n}\cdot\vec{\sigma}}=e^{-i\gamma_g\vec{n}\cdot\vec{\sigma}}.
\end{equation}
In contrast, for the two-loop CNGQGs within Path B, only $\phi_2$ and $\phi_5$ in Eq. (B4) need to be adjusted to $\beta_0-\gamma_g/2-\pi/2$, yielding the same evolution operator $U^2_c(\tau)=-e^{-i\frac{\gamma_g}{2}\vec{n}\cdot\vec{\sigma}}\cdot -e^{-i\frac{\gamma_g}{2}\vec{n}\cdot\vec{\sigma}}=e^{-i\gamma_g\vec{n}\cdot\vec{\sigma}}$. By adopting the same parameters $\alpha_0$, $\beta_0$ and $\gamma_g$ as specified in the main text, the corresponding geometric S and H gates can be obtained.

\section{The dynamical gates}
To construct the universal single-qubit dynamical gates, we start the Hamiltonian $\mathcal{H}_{d}(t)$ driven by a resonant microwave field, whose form is the same as that in Eq. (1), i.e.,
\begin{equation}\label{dynamicalHamiltonian}
\mathcal{H}_{d}(t)=\frac{\Omega_d(t)}{2}\left[\cos\phi_d(t)\sigma_x+\sin\phi_d(t)\sigma_y  \right].
\end{equation}
Here, we set $\phi_d(t)\equiv\phi_d$ being a constant to ensure these gates being pure dynamical gates. Thus, after an evolution period $\tau$, the evolution operator can be calculated as
\begin{eqnarray}
\begin{split}
U_d(\chi_d,\phi_d)&=e^{-i\int^{\tau}_0\mathcal{H}_d(t)dt}  \\
&=\left(
                                                \begin{array}{cc}
                                                  \cos(\frac{\chi_d}{2}) & -i\sin(\frac{\chi_d}{2})e^{-i\phi_d} \\
                                                  -i\sin(\frac{\chi_d}{2})e^{i\phi_d} & \cos(\frac{\chi_d}{2}) \\
                                                \end{array}
                                              \right),
\end{split}
\end{eqnarray}
with the pulse area $\chi_d=\int^{\tau}_0\Omega_d(t)dt$. By setting appropriate parameters, dynamical S and H gates can be constructed, respectively, as
\begin{eqnarray}
\begin{split}
S&=U_d\left(\frac{\pi}{2},\pi\right)U_d\left(\frac{\pi}{2},\frac{3\pi}{2}\right)U_d\left(\frac{\pi}{2},0\right),  \\
H&=U_d\left(\pi,0\right)U_d\left(\frac{\pi}{2},\frac{\pi}{2}\right).
\end{split}
\end{eqnarray}

\end{document}